\documentclass{elsart}

\usepackage{graphics}
\usepackage{graphicx}

\usepackage{amssymb}

\begin{document}

\begin{frontmatter}

\title{Echos of the liquid-gas phase transition in multifragmentation}
\author[gsi,ifin]{Al. H. Raduta},
\author[gsi,ifin]{Ad. R. Raduta}

\address[gsi]{GSI, D-64220 Darmstadt, Germany,}
\address[ifin]{NIPNE, RO-76900 Bucharest, Romania}

\begin{abstract}
  A general discussion is made concerning the ways in which one can
  get signatures about a possible liquid-gas phase transition in
  nuclear matter. Microcanonical temperature, heat capacity and second
  order derivative of the entropy versus energy formulas have been
  deduced in a general case. These formulas are {\em exact}, simply
  applicable and do not depend on any model assumption.  Therefore,
  they are suitable to be applied on experimental data. The formulas
  are tested in various situations. It is evidenced that when the
  freeze-out constraint is of fluctuating volume type the deduced
  (heat capacity and second order derivative of the entropy versus
  energy) formulas will prompt the spinodal region through specific
  signals.  Finally, the same microcanonical formulas are deduced for
  the case when an incomplete number of fragments per event are
  available. These formulas could overcome the freeze-out backtracking
  deficiencies.
\end{abstract}
\end{frontmatter}

\section{Introduction}

From a long time it is believed that, due to the van der Waals nature
of the nucleon-nucleon interaction, nuclear matter is likely to
exhibit a liquid-gas phase transition \cite{bertsch}.  While such a
phase transition was predicted within various models, from the
experimental point of view so far things are rather inconclusive. The
reasons are related to both the inherent difficulties of backtracking
the freeze-out stage and the lack of any rigorous mathematical
apparatus permitting the unambiguous determination of a liquid-gas
phase transition when it exists in experimental data.  Some years ago,
the general conviction was that the experimental measurement of the
nuclear caloric curve could give a definite answer to this problem.
Neglecting the weaknesses of each of the ``thermometers'' tested with
that occasion, there is one fundamental barrier preventing the
inference of a first order phase transition from any experimentally
obtained caloric curve even with back-bending shape: The
experimentally obtained microcanonical systems follow an {\em unknown}
path in the excitation energy ($E$)- freeze-out volume ($V$) plane
\cite{chomaz_prl}.  {\em Even} if the experimental path intersects the
coexistence region the phase transition may remain unrevealed in the
caloric curve \cite{chomaz_prl}.  For overcoming that situation it was
proposed that the heat capacity being a quantity {\em independent} on
the experimental path but depending on the {\em local} freeze-out
constrain could give precise information concerning the occurrence of
the transition \cite{chomaz_npa,chomaz_prl}. More exactly, {\em if}
the experimental path is intersecting the spinodal region of the
system's phase diagram, this would be signaled by a negative value of
the heat capacity.  An analytical heat capacity formula was proposed
with that occasion and was subsequently employed in investigating the
experimental data obtained by the MULTICS-MINIBALL collaboration
occasion with which negative branches of the heat capacity curves were
identified \cite{D'Agostino}. The formula was further applied on the
INDRA data with the same success \cite{D'Ago+indra}.  However, it is
worth noticing that though formally correctly deduced, this formula is
based on unevaluated terms (such as the microcanonical temperature
($T$) or the heat capacity ($C_1$) corresponding to a partial energy
of the system ($E_1$)) which further imply a chain of assumptions and
approximations \footnote{For example in Ref. \cite{D'Agostino} the
  temperature estimator does not correspond to the microcanonical one,
  is based on the assumption of thermal equilibrium between fragment
  internal and external degrees of freedom and, moreover, depends on
  additional level density parameters; furthermore, $C_1$ is deduced
  from the $T(\left<E_1\right>)$ dependence which results in even
  larger deviations.}  resulting in (unpredicted) deviations from the
exact results.

The above matters motivated us to draw the attention in the present
paper on an alternate way to calculate the microcanonical heat
capacity of nuclear systems proposed in Ref. \cite{c}. As will be
further shown this method is {\em exact}, model independent, simply
applicable and does not depend on any unevaluated term.  The
present paper is structured as follows: Section 2 gives a brief review
of the model used for illustrating the paper's ideas. Section 3
investigates on the possible signals one can (experimentally) get
about a possible liquid-gas phase transition. Microcanonical formulas
for temperature, heat capacity and second-order derivative of the
entropy versus energy are deduced for various conservation options in
section 4. Examples for the functioning of the above-mentioned formulas
are given in section 5. The same quantities are deduced in section 6
for the case in which only a percent of the total number of fragments
are available per event. Conclusions are drawn in section 7.
 
\section{Model review}

For testing the paper's argumentation we use the sharp microcanonical
multifragmentation model proposed in Ref.  \cite{model}. The model
concerns the disassembly of a statistically equilibrated nuclear
source $(A,Z,E,V)$ (i.e. the source is defined by the parameters: mass
number, atomic number, excitation energy, and freeze-out volume
respectively); its basic hypothesis is equal probability between all
configurations $C:\{A_i,~Z_i,~\epsilon_i,~{\bf r}_i,~{\bf
  p}_i,~~i=1,\dots,N\}$ (the mass number, the atomic number, the
excitation energy, the position and the momentum of each fragment $i$
of the configuration $C$, composed of $N$ fragments) which are subject
to standard microcanonical constraints: $\sum_i A_i=A$, $\sum_i
Z_i=Z$, $\sum_i {\bf p}_i=0$, $\sum_i {\bf r}_i\times{\bf p}_i=0$, $E$
- constant; integration over the fragments' momenta can be
analytically performed in the expression of the total number of
states, then one works in the smaller configuration space:
$C':\{A_i,~Z_i,~\epsilon_i,~{\bf r}_i,~~i=1,\dots,N\}$; a
Metropolis-type simulation is employed for determining the average
value of any system observable. Function of the desired conservation
restriction (${\mathcal R}$) level ($n$) one can write:
\begin{eqnarray}
W_{C'}\propto I_p=
\int\prod_{i=1}^N {\rm d} {\bf p}_i~{\mathcal R}(n) 
        =\frac{2\pi \prod_i m_i^{3/2}}{\Gamma\left(\frac32(N-n)\right)}
        f(n)
        \left(2\pi K
        \right)^{\frac32(N-n)-1},
        \label{eq:int}
\end{eqnarray}
with $n=0,1,2$; ${\mathcal R}(0)=\delta\left(H-E\right)$, ${\mathcal
  R}(1)={\mathcal R}(0)~\delta\left(\sum_i {\bf p}_i\right)$,
${\mathcal R}(2)={\mathcal R}(1)~\delta\left(\sum_i {\bf r}_i \times
  {\bf p}_i\right)$; $f(0)=1, f(1)=1/(\sum_i m_i)^{3/2}$ and
$f(2)=f(1)/\sqrt{\det {\bf I}}$, where ${\bf I}$ is the inertial
tensor of the system. Here $H$ denotes the system's Hamiltonian:
$H=\sum_i p_i^2/(2 m_i)+\sum_{i<j}V_{ij}+\sum_i\epsilon_i-\sum_i B_i$
and $K=E-\sum_{i<j}V_{ij}-\sum_i \epsilon_i+\sum_i B_i$ ($V_{ij}$ is
the Coulomb interaction between fragments $i$ and $j$; $\epsilon_i$
and $B_i$ are respectively the internal excitation and the binding
energy of fragment $i$). In the standard version of the model
fragments are idealized as nonoverlapping hard spheres placed into a
spherical recipient; intersection between fragments and the
``recipient's'' wall is also forbidden. We call this freeze-out
hypothesis (i).  In a simplified version, (ii), one can further
perform an approximate integration over the position variables as in
\cite{prl} and get the extra factor: $V_{free}=\prod_{i=1}^N
\left(V-i~V_0/N\right)$ in the expression of the statistical weight of
a configuration $C'':\{A_i,~Z_i,~\epsilon_i,~~i=1,\dots,N\}$ (i.e.
$W_{C''}=W_{C'} V_{free}$), where $V_0$ stands for the volume of the
nuclear system at normal nuclear matter density. In this version of
the model the Wigner Seitz approximation is used for the interfragment
Coulomb interaction. When angular momentum conservation is also
considered, $\det {\bf I}$ is approximated by the quantity
corresponding of a sphere of volume $V$ uniformly filled with nuclear
matter of density $(V_0/V) \rho_0$. The latter version is
computationally faster and provides a different freeze-out perspective
(i.e. spherical hard-core interaction is missing).

\section{Echos of the liquid-gas phase transition}

Which are the echos one can experimentally get from a liquid-gas phase
transition? For answering this question one should first define the
liquid-gas phase transition in {\em finite} {\em extensive} or {\em
  nonextensive} systems. This matter was recently addressed in
\cite{prl}. There, it was shown that the techniques usually employed
for very large systems work for the case of finite extensive or
nonextensive systems as well. Namely, it was shown that curves like
$Y(X)|_{Y'}$ (here $X$ is an extensive variable, $Y$ is the $X$'s
conjugate and $Y'$ is another intensive variable) have the property of
revealing the phase transition (even in small and nonextensive systems)
by bending backwards. The microcanonical isobaric caloric curves or
isothermal pressure versus volume curves are natural examples for the
physical situation under study. The probability distributions of an
isobaric canonical ensemble proved to be a precious tool for
identifying the phase transition when it exists (by exhibiting a
double peaked structure) and for constructing the corresponding
$T(E)|_P$ and $P(V)|_T$ curves. Reasons lay in the simple expression
of the probability of a state with excitation energy $E$ and volume
$V$ in a constant pressure canonical ensemble characterized by the
parameters $P$ (canonical pressure) and $\beta$ (inverse of the
canonical temperature):
\begin{eqnarray}
{\mathcal P}(E,V)=\frac{W(E,V)~e^{-\beta E- \beta P V}}{Z_{can}(\beta,P)},
\label{eq:p(e,v)}
\end{eqnarray}
leading to the expression of the microcanonical entropy ($S=\ln W(E,V)$):
\begin{eqnarray}
S(E,V)=\ln {\mathcal P} (E,V)+\beta E +\beta P V + \ln Z_{can} (\beta, P),
\label{eq:s(e,v)}
\end{eqnarray}
where $Z_{can}(\beta,P)$ represents the constant pressure canonical
partition function. Starting from their definitions
($T_{\mu}=(\partial S(E,V)/ \partial E)^{-1}$, $P_{\mu}=T_{\mu}
\partial S(E,V)/ \partial V$), one can easily obtain the
microcanonical temperature and pressure:
\begin{eqnarray}
T_{\mu}&=&\left(\beta+\frac{\partial \ln {\mathcal P} (E,V)}{\partial E} \right)^{-1},
\nonumber \\
P_{\mu}&=&T_{\mu} \left( \beta P + \frac{\partial \ln {\mathcal P}
    (E,V)}{\partial V} \right).
\label{eq:tp}
\end{eqnarray}
One can then follow the method proposed in \cite{prl} and follow a
$P_{\mu}=P$ or a $T_{\mu}=1/\beta$ path in the $(E,V)$ plane by using
the $(P_{\mu},T_{\mu})$ pairs evaluated (by means of eq.
(\ref{eq:tp})) in each $(E,V)$ point with sufficiently collected
events from the probability distribution of a canonical ensemble at
constant pressure (one can simulate this ensemble within the present
model by multiplying the statistical weight of a configuration by
$e^{-\beta E -\beta P V}$ and letting both energy and freeze-out volume
to fluctuate). The corresponding isobaric and isothermal paths are
plotted in Fig. 1 (upper-right panel) for a system with $A=200$,
$Z=82$ for which Coulomb interaction is switched to zero. The
corresponding constant pressure canonical probability distribution is
represented by a contour plot (in log scale) in the same picture. The
double peaked structure points the presence of a first order phase
transition. Both isothermal and isobaric paths are intersecting the
two peaks of the distribution and also a third saddle-like point. The
corresponding $T(E)|_P$ and $P(V)|_T$ curves are plotted in the lateral
panels with thick lines. They both present a backbending region
indicating the phase transition.
\begin{figure}
  \begin{center}
    \includegraphics[height=10cm]{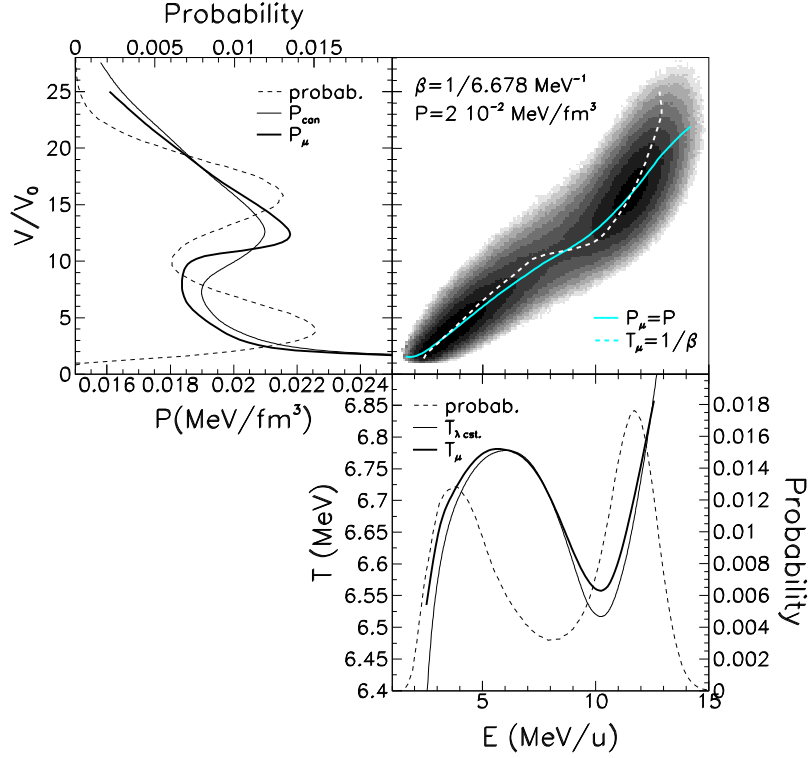}
    \caption
    {
      Right-upper panel: probability distribution (in log scale) of
      a canonical ensemble at constant pressure corresponding to the
      nucleus (200,82) where Coulomb interaction has been switched
      off. The canonical parameters $\beta$ and $P$ are labeled on the
      picture. The thick lines correspond to microcanonical 
      isobaric and isothermal paths corresponding to $P_{\mu}=P$
      and $T_{\mu}=1/\beta$. The lateral panels correspond to
      projections (dashed lines) 
      of the probability distributions on $V/V_0$ (left) and
      $E$ (down) axes. Canonical pressure (left panel) and
      constant-$\lambda$ temperature are figured with thin lines. The
      corresponding microcanonical curves are plotted with thick
      lines. (Here and in the rest of the figures $E$ stands for
      excitation energy.)
    }
    \label{fig:1}
  \end{center}
\end{figure}

While such constructions can be simply performed in theoretical
models, experimentally one can usually only access microcanonical
ensembles (due to the sorting of events in bins of $E$). Moreover, it
is widely believed that the events corresponding to a microcanonical
ensemble at a given $E$ have a fluctuating volume at freeze-out -
somewhat similar to the case of a volume constraint $e^{-\lambda V}$
proposed in Ref. \cite{chomaz_prl} (where $\lambda$ is the Lagrange
multiplier corresponding to the volume observable). As clearly
resulting from Fig. 1, the constant $\lambda$ microcanonical ensemble
corresponds to the folding on the $V$ axis of the constant pressure
canonical ensemble. Indeed, replacing in eq. (\ref{eq:p(e,v)}) $\beta
P $ with $\lambda$, and making the integration over $V$ one gets:
\begin{eqnarray}
{\mathcal P}^{\lambda}(E)=
\frac{W^{\lambda}(E)~e^{-\beta E}}{Z_{can}(\beta,P)},
\label{eq:p(e)_lambda}
\end{eqnarray}
and then the expression of the microcanonical temperature:
\begin{eqnarray}
T^{\lambda}_{\mu}&=&\left(\beta+\frac{\partial \ln {\mathcal P}^{\lambda}
    (E)}{\partial E} \right)^{-1}.
\label{eq:tp_lambda}
\end{eqnarray}
The dependence of the last two quantities on $E$ is represented in
Fig. 1 (lower panel). Thus, ${\mathcal P}^{\lambda}(E)$ is a
projection on the $E$ axis of the ${\mathcal P}(E,V)$ distribution.
Therefore, a double peaked structure in ${\mathcal P}^{\lambda}(E)$ is
{\em necessarily} related to a double peaked structure in ${\mathcal
  P}(E,V)$, and thus, is {\em reflecting} a first order liquid-gas
phase transition  
as defined in Refs. \cite{chomaz_pre,prl}.

Of course, the same reasoning may be done for the projection on the
$V$ axis. The corresponding curves are plotted in Fig. 1 (left
panel). However, the relevance of this latter projection for the
experimental data is conditioned by a good event by event resolution in
freeze-out volume which was not achieved so far.

\section{Microcanonical formulas}

Let us further deduce the {\em exact} expressions for
microcanonical temperature and heat capacity. The microcanonical
density of states corresponding to a total energy of the system $E$
writes:
\begin{eqnarray}
W(E)=\sum_{C'}W_{C'}
\label{eq:w(e)}
\end{eqnarray}
where 
\begin{eqnarray}
  W_{C'}=\frac1{N!}
  \left(
    \prod_{i=1}^N \frac{\rho_i(\epsilon_i)}{h^3}
  \right)
  F({\mathcal P}) f(n)
  \left[2\pi (E-{\mathcal E})\right]^{\frac32(N-n)-1}
\label{eq:w(c')}
\end{eqnarray}
and 
\begin{eqnarray}
  \sum_{C'}()\equiv
  \sum_{N=1}^A \int\!\!\!\!\!\!\!\sum{\rm d}{\mathcal E}()\equiv
  \sum_{N=1}^A\prod_{i=1}^N 
  \left(
    \sum_{A_i,Z_i}\int {\rm d} {\bf r}_i \int {\rm d} \epsilon_i 
  \right)()
  \label{eq:sum_c}
\end{eqnarray}                                
Here $F({\mathcal P})$ is the factor in front of $f(n)$ from eq.
(\ref{eq:int}) (${\mathcal P}$ is a generic notation for fragment
partition) and ${\mathcal E}\equiv\sum_{i<j}V_{ij}+\sum_i
\epsilon_i-\sum_i B_i$, the expression (\ref{eq:sum_c}) being formal.
Then,
eq. (\ref{eq:w(e)}) may be rewritten as:
\begin{eqnarray}
  W(E)=\sum_{N=1}^A\int\!\!\!\!\!\!\!\sum {\rm d}{\mathcal E}{\mathcal F}(C')
  \left(E-{\mathcal E}\right)^{\frac32(N-n)-1}.
  \label{eq:W(E)}
\end{eqnarray}
The microcanonical temperature writes $T^{-1}=\partial S/\partial E$
with $S=\ln W(E)$. Which means:
\begin{eqnarray}
  T^{-1}=\frac1{W(E)}\frac{\partial W(E)}{\partial E}
\end{eqnarray}
{\em If} the limits of the (formal) integral over ${\mathcal E}$ are
{\em not} depending on $E$, then one can simply make the derivative
versus $E$ {\em inside} the integral from eq. (\ref{eq:W(E)}):
\begin{eqnarray}
   T^{-1}=\frac1{W(E)}
   \sum_{C'} W_{C'}
   \left(\frac{\frac32(N-n)-1}{E-{\mathcal E}}\right)
   =\left<\frac{\frac32(N-n)-1}K\right>,
\label{eq:T}
\end{eqnarray}
where we used the implication: $K=E-{\mathcal E}$ and the
notation$\left<~\right>$ for the average over the ensemble's states.
The heat capacity of the system is by definition: $C^{-1}=-T^2 
\left(\partial^2 S/\partial E^2\right)$. Using the same arguments as
those from the deduction of $T$ one obtains:
\begin{eqnarray}
  C^{-1}=1-T^2
    \frac1{W(E)}\frac{\partial^2 W(E)}{\partial E^2},
\end{eqnarray}
which implies:
\begin{eqnarray}
  C^{-1}=1-T^2\left<
    \frac{\left[\frac32(N-n)-1\right]
      \left[\frac32(N-n)-2\right]}{K^2}
  \right>.
\label{eq:C}
\end{eqnarray}
Alternatively, one can (similarly) evaluate the second order derivative of
the system's entropy versus $E$:
\begin{eqnarray}
  \frac{\partial^2 S}{\partial E^2}=
  \left<
    \frac{\left[\frac32(N-n)-1\right]\left[\frac32(N-n)-2\right]}{K^2}
  \right>-
  \left<
    \frac{\frac32(N-n)-1}K
  \right>^2.
\label{eq:d2sde2}
\end{eqnarray}
In these terms, the system's spinodal region will be pointed by
positive values of the above quantity.  One can easily check that for
$n=1$ (i.e. energy and total momentum are the conserved quantities),
eq. (\ref{eq:C}) is similar with the one deduced in Ref.  \cite{c}.
The microcanonical formulas given by eqs. (\ref{eq:T}), (\ref{eq:C})
and (\ref{eq:d2sde2}) are {\em universally} applicable to {\em any}
system for which the ``external'' degrees of freedom can be treated
classically irrespective to other specificities of the system.
Indeed, the integral over the momenta variables (with the various
possible conservation options) given in eq.  (\ref{eq:int}), from
which the microcanonical $T$, $C$ and $\partial^2 S/\partial E^2$
formulas were further deduced is valid for any classical $N$ particle
system. This is the case of nuclear multifragmentation as well: it is
widely accepted that, since at freeze-out the system is rather dilute,
a classical treatment of the fragments' motion is appropriate. These
features make eqs. (\ref{eq:T}), (\ref{eq:C}) and (\ref{eq:d2sde2})
suitable to be applied for experimental data: they are {\em
  independent} on specific model assumptions (i.e. freeze-out
hypothesis, internal excitation treatment - level density,
interfragment interactions, etc.), and moreover, they only depend on
two parameters ($K$ and $N$) which have to be estimated for the
freeze-out stage.

\begin{figure}
\begin{center}
\includegraphics[height=11cm]{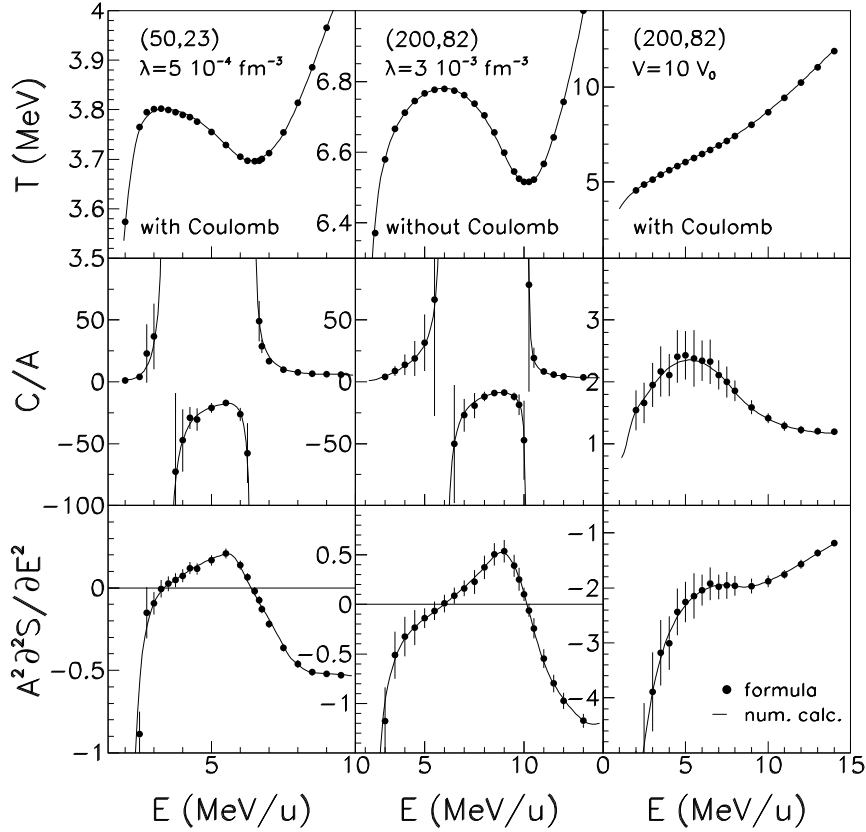}
\caption{Comparison between the results of the microcanonical formulas
  (for temperature, heat capacity and second order derivative of the
  entropy)
  and the corresponding curves deduced from the probability
  distributions of a canonical ensemble with a given freeze-out
  constraint (i.e. constant volume or constant $\lambda$). Each
  column correspond to the case labeled on it.
}
\label{fig:2}
\end{center}
\end{figure}

\section{Examples}

While the deduction of the above formulas guarantees their accuracy,
some examples of their functioning on concrete cases are further
discussed. As an independent test, the results will be compared with
the temperature/heat capacity computed using the method synthesized by
eq.  (\ref{eq:tp_lambda}). More exactly, the caloric test curves are
computed using eq. (\ref{eq:tp_lambda}) from the probability
distributions of a canonical ensemble with a given freeze-out
constraint and the heat capacity and $\partial^2 S/\partial E^2$
curves are farther evaluated from the obtained caloric curves
($C=\partial E/\partial T$, $\partial^2 S/\partial E^2=(-1/T^2)
\partial T/\partial E$).  Calculations are performed with both
versions of the model ((i) and (ii)), for two different freeze-out
constraints: constant $\lambda$ and constant volume, for the case in
which the Coulomb interaction is present and the one in which it is
switched off.  (This variety of examples constitute a further test for
the universality of the deduced formulas.) The considered conservation
``level'' is $n=2$, i.e.  the total energy, the momentum and the
angular momentum are conserved quantities. Results corresponding to
the cases: (50,23) nucleus with constant $\lambda=5~10^{-4}$ fm$^{-3}$
with Coulomb interaction included, version (ii) of the model,
(200,82), $\lambda=3~10^{-3}$ fm$^{-3}$, without Coulomb interaction,
version (ii) of the model and the same nucleus with Coulomb
interaction included, but a constant volume constraint: $V=10~V_0$,
version (i) of the model are presented in Fig. 2 (first, middle and
third column respectively). In the first two cases, backbendings of
the caloric curves reflected in negative branches of the heat capacity
curves and positive regions in the second order derivative of the
entropy versus $E$ - all related with a first order phase transition
can be observed. In the third case (constant volume constraint) the
caloric curve has a monotonic increase with a ``plateau-like'' region
(where the curve slope is smaller, but positive) which is reflected
in a positive $C$ (with a peak in the plateau-like region) and a
negative $\partial^2 S/\partial E^2$. (As will be further seen, {\em
  one cannot} draw any conclusion from the third case concerning the
intersection of the coexistence region since a constant volume path is
{\em not necessarily} the order parameter of the system.) A very good
agreement between the curves calculated using the probability
distribution of a canonical ensemble and the points evaluated by means
of the microcanonical formulas is to be noticed. The calculated
microcanonical heat capacity and second derivative of the entropy
versus excitation energy points are clearly revealing the spinodal
region. While the calculated heat capacity points have large error
bars near the border of the spinodal region due to the asymptotic
behavior of the heat capacity, $\partial^2 S/\partial E^2$ has small
error bars even in the border region - thus being an even more robust
quantity for identifying the spinodal region.
\begin{figure}
\begin{center}
\includegraphics[height=9cm]{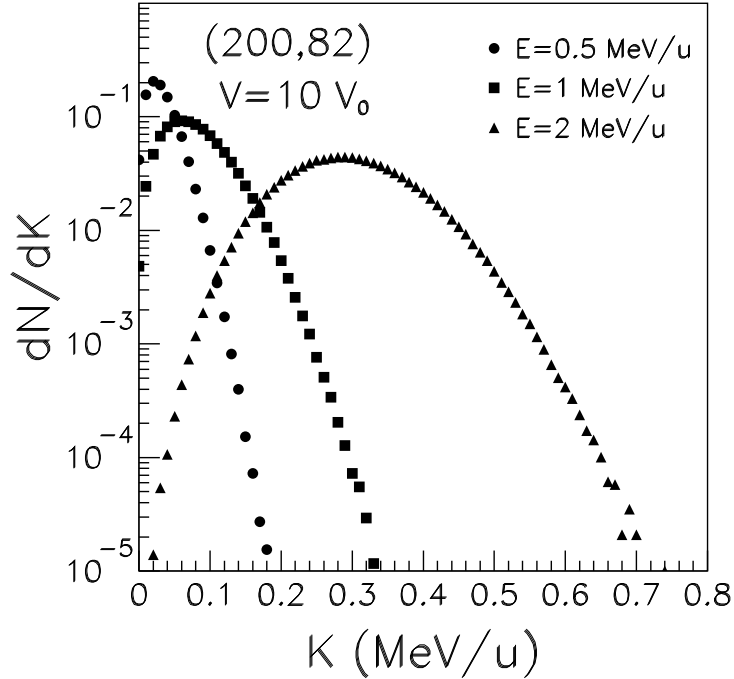}
\caption{Kinetic energy distribution corresponding to the source
  nucleus (200,82) with the freeze-out volume $V=10~V_0$ with the
  Coulomb interaction included, for three excitation energies (0.5, 1
  and 2 MeV/u).
}
\label{fig:3}
\end{center}
\end{figure}

Some details concerning the applicability of the microcanonical
formulas have to mentioned. As mentioned earlier, the validity of
these formulas is restricted by the condition that the integration
domains in eq. (\ref{eq:W(E)}) over the quantity ${\mathcal E}$ {\em not
  to be restricted} by the total energy of the system $E$. One should
thus limit the cases on which this formula is applied to those obeying
the above condition. But how to distinguish those ``forbidden'' points
from the rest? A restriction by $E$ of the upper limit of the integral
over ${\mathcal E}$ has to be clearly pointed by a ``cut'' in the right
part of ${\mathcal E}$ distribution. Since $K=E-{\mathcal E}$, a similar
``cut'' (corresponding to $K=0$) has to be noticed in the left part of
the $K$ probability distribution. In other words, when the probability
that $K=0$ is significant (i.e. a cut clearly affects the $K$
distribution) then deviations from the microcanonical formulas should
occur. These ``cuts'' (usually) occur at very low energies, so in a
region unimportant for the system's thermodynamics. All the points
represented in Fig. 2 have been checked against the cut in the $K$
probability distribution. For illustrating this idea we take the case
(200,82), with Coulomb, $V=10~V_0$. The probability distributions of
$K$ corresponding to the $E=$0.5, 1 and 2 MeV/u cases are illustrated
in Fig. 3. While the $K$ probability distributions corresponding to
$E=$0.5 and 1 MeV/u are intersecting the $K=0$ axis, (giving thus
a non-negligible probability for $K=0$), the probability distribution
corresponding to the $E=2$ MeV/u case is not intersecting the $K=0$
vertical axis. This fixes the lowest excitation energy limit for which
the microcanonical formula works to some value between 1 and 2 MeV/u. 

\begin{figure}
\begin{center}
\includegraphics[height=12cm]{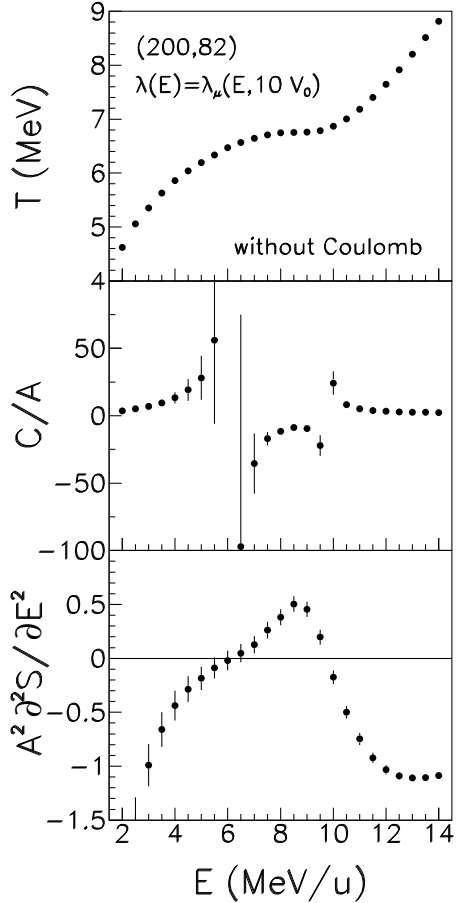}
\caption{
  Caloric curves, heat capacity curves and second order derivative of
  the entropy curves evaluated with the microcanonical formulas for
  the case of the nucleus (200,82), without Coulomb interaction and a
  freeze-out constrain corresponding to a path in $\lambda$ given by:
  $\lambda(E)=\lambda_{\mu}(E, 10~V_0)$.
}
\label{fig:4}
\end{center}
\end{figure}

{
  The crossing of the spinodal region by a (experimental) path in
  the (E,V) plane {\em may not} be reflected as a backbending in a
  caloric curve. However, if the volume constraint is of the
  $e^{-\lambda V}$ type the spinodal region {\em is likely} to be
  revealed by a negative branch in the heat capacity}.  This point was
nicely argued in Ref. \cite{chomaz_prl} and can be evidenced by means of
the present microcanonical formulas as a further test for their
abilities. To this aim, we chose $\lambda$ as the microcanonical
``$\lambda$'' parameter ($\lambda_{\mu}$) corresponding to a constant
volume path ($V=10~V_0$) in the (E,V) plane. I.e.:
\begin{eqnarray}
  \lambda(E)=\lambda_{\mu}(E,10~V_0).
  \label{eq:lambda(E)}
\end{eqnarray}
The microcanonical $\lambda$ can be simply obtained from eq. (\ref{eq:tp}):
\begin{eqnarray}
  \lambda_{\mu}=\lambda+\frac{\partial \ln {\mathcal P}
    (E,V)}{\partial V}.
\end{eqnarray}
As illustrated in Fig. 4 this path appears to intersect the system's
spinodal region without inducing any backbending in the caloric curve
for the case of the system (200,82) without Coulomb interaction, (ii)
version of the model. Indeed, while the caloric curve has a
monotonical increase, the corresponding heat capacity presents a
negative branch and the second order derivative of the entropy a
positive region. The negative region in the heat capacity and the
positive region in the second order derivative of the entropy versus
energy are evidencing a ``spinodal line'' of the system corresponding
to the volume $V=10~V_0$ in the system's phase diagram in $(E,V)$
representation.  The limits $E$ of this region are {\em approximately}
those delimited by the above mentioned signals.

\section{Fewer fragments}

Let's now assume that experimentally only a (small) number $N_1$
($<N$) of fragments corresponding to a given event are detected. Can
one still get the desired thermodynamical information about the
system? This question is addressed in the present section.  Let us
turn back to eq. (\ref{eq:int}). For the highest conservation level
($n=2$) we can write:
\begin{eqnarray}
 I_p&=&\int\prod_{i=1}^N {\rm d} {\bf p}_i~
 \delta\left(\sum_i\frac{p_i^2}{2m_i}-K\right)
 \delta\left(\sum_i {\bf p}_i\right)
 \delta\left(\sum_i {\bf r}_i \times {\bf p}_i\right) \nonumber \\
 &=&\int\prod_{i=N_1+1}^{N} {\rm d} {\bf p}_i
 \int\prod_{i=1}^{N_1} {\rm d} {\bf p}_i~
 \delta\left(\sum_{i=1}^{N_1}\frac{p_i^2}{2m_i}-K_1\right)
 \delta\left(\sum_{i=1}^{N_1}{\bf p}_i-{\bf P}_1\right)
 \delta\left(\sum_{i=1}^{N_1} {\bf r}_i \times {\bf p}_i-{\bf L}_1\right),
\end{eqnarray}
where $K_1=E-{\mathcal E}-\sum_{i=N_1+1}^N p_i^2/(2m_i)$,
${\bf P}_1=-\sum_{i=N_1+1}^N {\bf p}_i=\sum_{i=1}^{N_1} {\bf p}_i$ and 
${\bf L}_1=-\sum_{i=N_1+1}^{N} {\bf r}_i \times {\bf p}_i=
\sum_{i=1}^{N_1} {\bf r}_i \times {\bf p}_i$. The
integration over the momenta of the first $N_1$ fragments can be
further performed:
\begin{eqnarray}
  I_p &\propto& \int\prod_{i=N_1+1}^{N} {\rm d} {\bf p}_i 
  \left(
    K_1-\frac{{\bf P}_1^2}{2M_1}-\frac12 {\bf L_1}^T{\bf I}_1^{-1}
    {\bf L_1}
  \right)^{\frac32(N-2)-1} \nonumber \\
  &\times& \theta\left(
   K_1-\frac{{\bf P}_1^2}{2M_1}-\frac12 {\bf L_1}^T {\bf I}_1^{-1}
  {\bf L_1} \right)
\end{eqnarray}
where ${\bf I}_1$ is the inertial tensor corresponding to the first
$N_1$ fragments at freeze-out, $M_1\equiv\sum_{i=1}^{N_1} m_i$
 and $\theta$ stands for the step
function. One can further deduce the thermodynamical quantities
obtained before similarly, starting from their statistical
definitions. In the general case (i.e a generic ``level'' of
conservation $n$) one obtains:
\begin{eqnarray}
  T^{-1}=\left[\frac32(N_1-n)-1\right]\left<\frac1{K_2}\right>,
\end{eqnarray}
\begin{eqnarray}
  C^{-1}=1-T^2\left[\frac32(N_1-n)-1\right]\left[\frac32(N_1-n)-2\right]
  \left<
    \frac1{K_2^2}
  \right>,
\end{eqnarray}
\begin{eqnarray}
  \frac{\partial^2 S}{\partial E^2}=
  \left[\frac32(N_1-n)-1\right]\left[\frac32(N_1-n)-2\right]
  \left<
    \frac1{K_2^2}
  \right> -
  \left[ \frac32(N_1-n)-1 \right]^2
  \left<
    \frac1{K_2}
  \right>^2,
\end{eqnarray}
where $K_2= K_1-{\mathcal K}(n)$ with ${\mathcal K}(0)=0$, ${\mathcal K}(1)={\bf
  P}_1^2/(2M_1)$ and ${\mathcal K}(2)={\mathcal K}(1)+{\bf L_1}^T {\bf
  I}_1^{-1}{\bf L_1}/2$. Following the same argumentation as in
section 4 one can deduce the criterion of validity of the above
formulas: the probability distribution of $K_2$ must {\em not}
intersect the $K_2=0$ vertical axis at a non-negligible value.  The
advantages of the above formulas are obvious: they only depend on one
parameter ($K_2$) and, more importantly, they can be used for
inferring information about the liquid-gas phase transition even when
one has {\em incomplete} information about the fragmentation events
(i.e. only $N_1$ fragments from the $N$ fragments of one fragmentation
event are detected). For a given microcanonical formula, depending on
the desired conservation level ($n$), the minimum allowed value of
$N_1$ is fixed by the condition that the factors containing $N_1$
be positive. Thus, for example, when only the total energy, $E$, is
conserved ($n=0$) one only needs {\em one} freeze-out fragment per event in
order to obtain the correct microcanonical temperature! 

\section{Summary}

In summary, the possibilities of obtaining information about a
possible liquid-gas phase transition from experimental heavy ion
collision data have been investigated. The connection between negative
heat capacity (or positive $\partial S^2/\partial E^2$) obtained in a
fluctuating volume ensemble and the liquid gas phase transition has
been discussed. It was shown that a negative value of the heat
capacity in a fluctuating volume (constant $\lambda$) ensemble is in
connection with a first order phase transition as defined in Ref.
\cite{prl}.  Microcanonical formulas for temperature, heat capacity
and second order derivative of the entropy versus energy have been
rigorously deduced for various conservation options (conserved energy;
conserved energy and momentum; conserved energy, momentum and angular
momentum).  These formulas were tested with an independent method for
obtaining the above mentioned quantities which extracts the
microcanonical information from the probability distributions of a
canonical ensemble with a given freeze-out volume constraint. An
excellent agreement between the two methods have been obtained for all
the considered curves ($T(E)$, $C(E)$, $\partial^2 S/ \partial E^2
(E)$) for various cases ((50,23), $\lambda=5~10^{-4}$ fm$^{-3}$ with
Coulomb, (200,82), $\lambda=3~10^{-3}$ fm$^{-3}$, without Coulomb,
(200,82), $V=10~V_0$ with Coulomb). In the first two cases,
backbendings in the caloric curves reflected in negative branches of
the heat capacity curves and positive regions in the $\partial^2 S/
\partial E^2 (E)$ curves have been obtained reflecting the occurrence
of a first order phase transition. In the third case no backbending in
the caloric curve or negative region in the heat capacity curve was
observed which {\em however} does not imply any conclusion about the
occurrence of a first order phase transition simply because a constant
volume path in the $(E,V)$ plane is {\em not necessarily} the system's
order parameter. Indeed, choosing for the nucleus (200,82) without
Coulomb a dependence of $\lambda$ on $E$ corresponding to the
microcanonical lambda with $V=10~V_0$
($\lambda(E)=\lambda_{\mu}(E,10~V_0)$) in an ensemble with fluctuating
volume one gets a well defined negative branch in the heat capacity
curve and a positive region of the second order derivative of the
entropy {\em in spite of} a monotonically increasing caloric curve.
The case of incomplete (experimental) information concerning the
fragmentation event is further discussed. Microcanonical formulas
corresponding to all conservation levels depending only on a given
number of fragments (smaller then the total number of fragments from
the given event) are deduced. While demanding some more kinematical
information about the detected fragments (only for the $n>0$ cases)
their utility is obvious. Thus, for example in the $n=0$ case one only
needs the kinetic energy of {\em one} freeze-out fragment per event in
order to deduce the correct microcanonical temperature.  The resulted
microcanonical formulas are model independent thus being applicable on
experimental data.  Moreover they do not depend on any unevaluated
term (such as heat capacity of a partial energy or the level density
parameters as is the case in Ref.\cite{D'Agostino}).  Of course, the
success of these formulas depends on the accuracy of backtracking the
information about the primary fragments. Here the microcanonical
formulas based on partial break-up information may be very effective
since some particles (such as freeze-out neutrons) carry at the
asymptotic stage the untouched information about the freeze-out.

The authors thank the Alexander von Humboldt Foundation for supporting 
this work.

\end{document}